\documentclass[
reprint,
superscriptaddress,
amsmath,amssymb,
aps,
prb,
prstab,
floatfix,
longbibliography
]{revtex4-2}

\usepackage[caption=false]{subfig}
\usepackage{multirow}
\usepackage{graphicx}
\usepackage{dcolumn}
\usepackage{bm}
\usepackage{threeparttable}
\usepackage{color}
\usepackage[dvipsnames]{xcolor}
\usepackage{chemformula}
\usepackage{xspace}
\usepackage{orcidlink}
\usepackage{hyperref}
\usepackage{booktabs}
\usepackage{makecell}
\usepackage{mathtools}
\hypersetup{colorlinks=true,linkcolor=blue,filecolor=blue,urlcolor=blue,citecolor=blue,pdftitle = {Terahertz spin-orbit torque driven spin dynamics in the insulator antiferromagnet Cr$_{2}$O$_{3}$}, pdfauthor = {R.~M.~Dubrovin}}

\newcommand{\CrO}{$\mathrm{Cr}_{2}\mathrm{O}_{3}$}

\bibpunct{[}{]}{,}{n}{}{}

\definecolor{newtext}{RGB}{0, 0, 0}
\definecolor{newtext2}{RGB}{0, 0, 0}
\definecolor{redtext}{RGB}{0, 0, 0}

\begin{document}

\preprint{APS/123-QED}


\title{Terahertz spin-orbit torque as a drive of spin dynamics\\ in insulating antiferromagnet Cr$_{2}$O$_{3}$}

\author{R.~M.~Dubrovin\,\orcidlink{0000-0002-7235-7805}}
\email{dubrovin@mail.ioffe.ru}
\affiliation{Ioffe Institute, Russian Academy of Sciences, 194021 St.\,Petersburg, Russia}

\author{Z.~V.~Gareeva\,\orcidlink{0000-0003-3690-6720}}
\affiliation{Institute of Molecule and Crystal Physics, Ufa Federal Research Center,  Russian Academy of Sciences, 450075 Ufa, Russia}
\affiliation{Ufa University of Science and Technology, 450076 Ufa, Russia}

\author{A.~V.~Kimel\,\orcidlink{0000-0002-0709-042X}}
\affiliation{Institute for Molecules and Materials, Radboud University, 6525 AJ Nijmegen, The Netherlands}

\author{A.~K.~Zvezdin\,\orcidlink{0000-0002-6039-780X}}
\email{zvezdin.ak@phystech.edu}
\affiliation{New Spintronic Technologies LLC, 121205 Skolkovo, Moscow, Russia}
\affiliation{Prokhorov General Physics Institute, Russian Academy of Sciences, 119991 Moscow, Russia}
\affiliation{Amirkhanov Institute of Physics, Dagestan Federal Research Center, Russian Academy of Sciences, 367003 Makhachkala, Russia}

\date{\today}

\begin{abstract}

Contrary to conventional wisdom that spin dynamics induced by current are exclusive to metallic magnets, we theoretically predict that such phenomena can also be realized in magnetic insulators, specifically in the magnetoelectric antiferromagnet $\mathrm{Cr}_{2}\mathrm{O}_{3}$.
We reveal that the displacement current driven by the THz electric field is able to generate a N{\'e}el spin-orbit torque in this insulating system.
By introducing an alternative electric dipole order parameter, i.e., the antiferroelectric vector, arising from the dipole moment at $\mathrm{Cr}^{3+}$ sites, we combine symmetry analysis with a Lagrangian approach and uncover that the displacement current couples to the antiferromagnetic spins and enables ultrafast control of antiferromagnetic order.
The derived equations of motion show that this effect competes with the linear magnetoelectric response, offering an alternative pathway for manipulating antiferromagnetic order in insulators.
Our findings establish insulator antiferromagnets as a viable platform for electric field-driven antiferromagnetic spintronics and provide general design principles for non-metallic spin-orbit torque materials.

\end{abstract}

\maketitle

\section{Introduction}

After three decades of successful development, ferromagnetic spintronics has begun to face fundamental challenges~\cite{han2023coherent}, which can no longer be resolved by marginal improvements and requires a paradigm shift. 
Antiferromagnetic spintronics is seen as one of the most promising routes for further developments.
Simultaneously, antiferromagnetic spintronics is also a challenge, as \textcolor{newtext2}{collinear} antiferromagnets have no net magnetization and are weakly sensitive to external magnetic fields.
Finding the most efficient ways to control spins in antiferromagnets has already for many decades been a topic of research~\cite{takeuchi2025electrical,rimmler2025non,dal2024antiferromagnetic,yan2024review,baltz2024emerging}.
The ability to generate strong and nearly single cycle THz pulses of electromagnetic radiation has managed to become a game changer in the field.
Using the pulses, it was demonstrated that spin dynamics in antiferromagnets with no net magnetization can be launched by both THz magnetic~\cite{zvezdin1979dynamics,andreev1980symmetry,zvezdin1981new,satoh2010spin} and electric fields~\cite{bilyk2025control}.
Electric current is also able to drive spin dynamics in metallic antiferromagnets, e.g., $\mathrm{Mn}_{2}\mathrm{Au}$ and $\mathrm{CuMnAs}$, at low, near-zero frequencies, through the N{\'e}el spin-orbit torque~\cite{wadley2016electrical,bodnar2018writing,manchon2019current,troncoso2019antiferromagnetic,selzer2022current,kaspar2021quenching,freimuth2021laser,behovits2023terahertz,ross2024ultrafast,olejnik2024quench}.
The origin of this torque is the Rashba–Edelstein effect, in which an electric current leads to spin polarization with different signs on different antiferromagnetic sublattices.
A similar mechanism was anticipated for \CrO{} at doping into the semiconducting or metallic regime~\cite{thole2020concepts}.
While the prospect of switching at THz rates has been one of the key motivations for fundamental studies in antiferromagnetic spintronics, such switching has not been experimentally demonstrated \textcolor{newtext2}{even for metallic antiferromagnets}~\cite{behovits2023terahertz}.
This challenge has driven an intensive search for new mechanisms capable of exciting spins in antiferromagnets at THz rates, and in particular, of generating THz spin oscillations~\cite{behovits2023terahertz,dubrovin2025competition,cao2025nonlinear,feng2025intrinsic}.

Here, we present a theoretical study of the so far overlooked effect of THz electric fields on antiferromagnetic spins. 
The effect is similar to the N{\'e}el spin-orbit torque, while originating from displacement currents and thus quite possible in materials with a poor electric conductivity, such as \CrO.
Using this prototypical insulating magnetoelectric antiferromagnet, we theoretically demonstrate that even in the absence of conduction electrons when Ohm's currents can be neglected, control of antiferromagnetic spins is possible via coupling the antiferromagnetic order parameter to the displacement current induced by the terahertz electric field.
We have considered the crystal structure of \CrO{} 
and found electric dipole moments ordered antiparallel along the antiferromagnetic ordering axis on the sites of $\mathrm{Cr}^{3+}$ magnetic ions.
Furthermore, we have shown that from symmetry, it is necessary to include the electric dipole order parameter in this coupling.  
Employing a Lagrangian approach, we derived the equations of spin dynamics in which the spin-orbit torque and the linear magnetoelectric effect enter in a similar way. 
Thus, in addition to the earlier reported mechanisms to control spins in antiferromagnets using magnetic and electric fields or electric current, here we demonstrate control using displacement currents similar to the N{\'e}el spin-orbit torque.
The mechanism is expected to be especially strong in insulating antiferromagnets, where the conventional N{\'e}el spin-orbit torque has been neglected so far.
\textcolor{newtext2}{We note that, in metals, the displacement current is negligible compared to electric current because the electric field is screened by free charge careers, the frequency is much lower than the conductivity $\omega \ll \sigma$, which in the Gaussian system has the same units of s$^{-1}$, and the field does not penetrate deep inside.}

\section{Material}

\begin{figure}
\centering
\includegraphics[width=1\columnwidth]{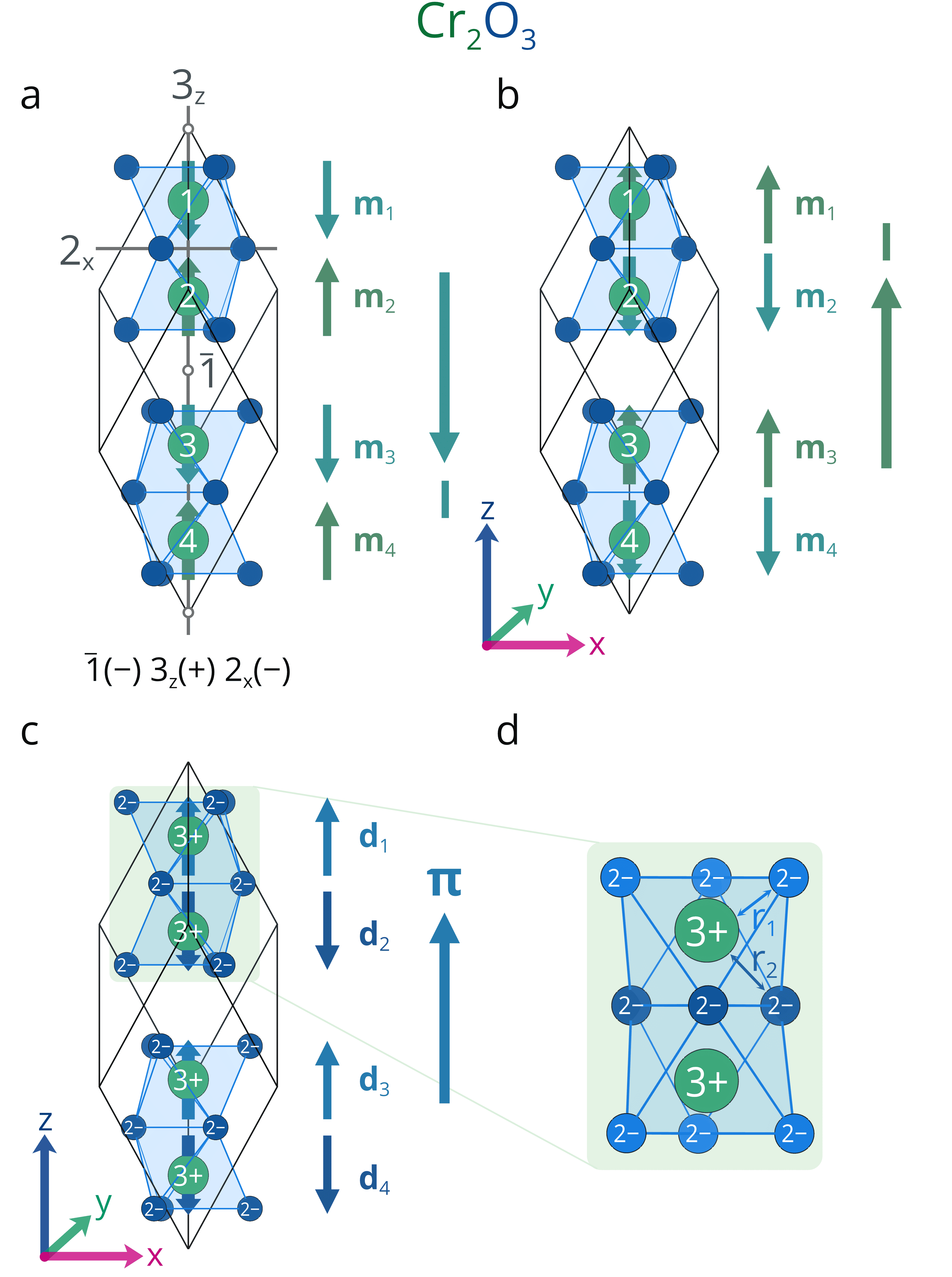}
\caption{\label{fig:structure}
Crystal and magnetic structures of antiferromanget \CrO{} with oppositely directed antiferromagnetic vectors (a)~$\mathbf{l}_{\downarrow}$ and (b)~$\mathbf{l}_{\uparrow}$.
Green arrows denote the orientation of magnetic moments $\mathbf{m}_{1}$--$\mathbf{m}_{4}$ of $\mathrm{Cr}^{3+}$ ions (labelled 1--4).
Positions of the symmetry elements $\overline{1}$, $3_{z}$, and $2_{x}$ in the unit cells are given in panel~(a).
(c)~Electric dipole moments $\mathbf{d}_{1}$--$\mathbf{d}_{4}$ (blue arrows) in the unit cell in the vicinity of magnetic $\mathrm{Cr}^{3+}$ ions.
The nominal charges of $\mathrm{Cr}^{3+}$ and $\mathrm{O}^{2-}$ ions in $e$ are given.
(d)~The nearest $\mathrm{O}^{2-}$ cations are located at two different distances $r_{1}$ and $r_{2}$ from the $\mathrm{Cr}^{3+}$ ions as marked in light and dark blue. 
}
\end{figure}

\CrO{} with corundum structure (space group $R\overline{3}c$, $Z=2$) is the prototypical magnetoelectric antiferromagnet~\cite{dzyaloshinskii1960magneto,astrov1960magnetoelectric,astrov1961magnetoelectric,rado1961observation}. 
Below the N{\'e}el temperature $T_{\mathrm{N}} = 307$\,K~\cite{volger1952anomalous,mcguire1956antiferromagnetism}, the magnetic moments of $\mathrm{Cr}^{3+}$ with the same value of magnetization $|\mathbf{m}_{1}| = |\mathbf{m}_{2}| = |\mathbf{m}_{3}| = |\mathbf{m}_{4}| = m_{0}$ are antiferromagnetically ordered along the $3_{z}$ axis, resulting in one of two types of antiferromagnetic ordering~\cite{brockhouse1953antiferromagnetic,corliss1965magnetic,bousquet2024sign}, as shown in Figs.~\ref{fig:structure}(a) for $\downarrow\uparrow\downarrow\uparrow$ and~\ref{fig:structure}(b) for $\uparrow\downarrow\uparrow\downarrow$. 
It is convenient to use the double-sublattice approximation in which codirectional magnetizations are replaced by two normalized magnetizations $\mathbf{m}_{\mathrm{A}} = \cfrac{\mathbf{m}_{1} + \mathbf{m}_{3}}{2 \, m_{0}}$ and $\mathbf{m}_{\mathrm{B}} = \cfrac{\mathbf{m}_{2} + \mathbf{m}_{4}}{2 \, m_{0}}$.
This allows us to introduce the net magnetization vector $\mathbf{m} = \cfrac{\mathbf{m}_{\mathrm{A}} + \mathbf{m}_{\mathrm{B}}}{2}$ and the antiferromagnetic N{\'e}el vector $\mathbf{l} = \cfrac{\mathbf{m}_{\mathrm{A}} - \mathbf{m}_{\mathrm{B}}}{2}$.
Further, we will use the Cartesian coordinate system in which $z \parallel \mathbf{l}$ as shown in Fig.~\ref{fig:structure}.

It is interesting to note that if we attribute nominal charges to atomic cores, e.g., $+3 \, e$ for $\mathrm{Cr}^{3+}$ and $-2 \, e$ for $\mathrm{O}^{2-}$, neglecting the Born corrections to the ionic charges, so that the electroneutrality of the \CrO{} unit cell is preserved, in the simple point charge model~\cite{gabbasova1991bifeo3,ederer2005influence,zvezdin2021multiferroic}, the nonzero electric dipole moments $\mathbf{d}_{i}$ oriented along the $z$ axis appear in the vicinity of $\mathrm{Cr}^{3+}$ ions $i$ = 1--4, as shown in Fig~\ref{fig:structure}(c).
\textcolor{newtext2}{The dipole moment at $i$th site is given by $\mathbf{d}_{i} = q \, \sum_{j=1}^{6} \mathbf{r}_{j}$, where $q$ is the anion charge and $\mathbf{r}_{j}$ is the radius vector from $i$th $\mathrm{Cr}^{3+}$ cation to the $j$th neighbour $\mathrm{O}^{2-}$ anion.
At the same time, the charge of $i$th $\mathrm{Cr}^{3+}$ cation does not contribute to the $\mathbf{d}_{i}$ at its own site.}
Thus, the dipole moments originate from the asymmetric arrangement of $\mathrm{O}^{2-}$ anions at two distinct distances $r_{1}$ and $r_{2}$ from the $\mathrm{Cr}^{3+}$ cation, as can be seen in detail in Fig.~\ref{fig:structure}(d), where equidistant oxygens are represented by color.

We emphasize that, according to Ref.~\cite{resta2007theory}, the dipole moment can be represented as $\mathbf{d} = \mathbf{d}_{\mathrm{ion}} + \mathbf{d}_{\mathrm{el}}$, where the ionic contribution $\mathbf{d}_{\mathrm{ion}}$ is estimated by the point charge model, while the electronic term $\mathbf{d}_{\mathrm{el}}$ accounting for the covalent bonding between the atoms is calculated using the Berry-phase approach.
\textcolor{newtext}{For simplicity, in our model we treat \CrO{} as a fully ionic crystal and thus neglect the electronic contribution $\mathbf{d}_{\mathrm{el}}$.} 
\textcolor{newtext}{Although it is known that \CrO{} has a magnetic-field-induced electronic polarization comparable to the ionic one~\cite{bousquet2011unexpectedly}, including it would quantitatively change the total polarization magnitude but not the qualitative picture of the electric ordering.}   
\textcolor{newtext}{Thus, analogous to the magnetic moments, these dipole moments are of equal magnitude $|\mathbf{d}_{1}| = |\mathbf{d}_{2}| = |\mathbf{d}_{3}| = |\mathbf{d}_{4}| = d_{0}$, and are arranged in such a way that the total dipole moment is zero for the unit cell.}
Note that such ordering of dipole moments is close to that observed in antiferroelectrics~\cite{kittel1951theory,tagantsev2013origin,wang2025type}.
Next, by analogy with antiferromagnets, we will use the double-sublattice approximation and define two normalized dipole moments $\mathbf{d}_{\mathrm{A}} = \cfrac{\mathbf{d}_{1} + \mathbf{d}_{3}}{2 \, d_{0}}$ and $\mathbf{d}_{\mathrm{B}} = \cfrac{\mathbf{d}_{2} + \mathbf{d}_{4}}{2 \, d_{0}}$, and the net polarization $\mathbf{p} = \cfrac{\mathbf{d}_{\mathrm{A}} + \mathbf{d}_{\mathrm{B}}}{2}$ and antiferroelectric vector $\bm{\pi} = \cfrac{\mathbf{d}_{\mathrm{A}} - \mathbf{d}_{\mathrm{B}}}{2}$.
\textcolor{newtext2}{The dipole moments for ionic crystals can be estimated experimentally by, for example, combining x-ray diffraction to determine ion positions with x-ray photoemission spectroscopy to determine their charges.}

\section{Results and Discussion}

\subsection{Symmetry analysis}

\begin{table}
\caption{\label{tab:generators} Permutation transformations of $\mathrm{Cr}^{3+}$ ions $1$--$4$, magnetic $\mathbf{m}$, $\mathbf{l}_{1}$--$\mathbf{l}_{3}$ and electric $\mathbf{p}$, $\bm{\pi}_{1}$--$\bm{\pi}_{3}$ basis vectors under the action of generators $\overline{1}$, $3_{z}$, and $2_{x}$ of the group for \CrO.}
\begin{ruledtabular}
\begin{tabular}{ccccccccccccc}
               & 1 & 2 & 3 & 4 & $\mathbf{m}$ & $\mathbf{l}_{1}$ & $\mathbf{l}_{2}$ & $\mathbf{l}_{3}$ & $\mathbf{p}$ & $\bm{\pi}_{1}$ & $\bm{\pi}_{2}$ & $\bm{\pi}_{3}$\\ \midrule
$\overline{1}$ & 4 & 3 & 2 & 1 & $\mathbf{m}$ &             $\mathbf{l}_{1}$ & $\mathllap{-}\mathbf{l}_{2}$ & $\mathllap{-}\mathbf{l}_{3}$ & $\mathllap{-}\mathbf{p}$ & $\mathllap{-}\bm{\pi}_{1}$ & $\bm{\pi}_{2}$ &            $\bm{\pi}_{3}$\\
$3_{z}$        & 1 & 2 & 3 & 4 & $\mathbf{m}$ &             $\mathbf{l}_{1}$ &             $\mathbf{l}_{2}$ &             $\mathbf{l}_{3}$ &             $\mathbf{p}$ &             $\bm{\pi}_{1}$ & $\bm{\pi}_{2}$ &             $\bm{\pi}_{3}$\\
$2_{x}$        & 2 & 1 & 4 & 3 & $\mathbf{m}$ & $\mathllap{-}\mathbf{l}_{1}$ & $\mathllap{-}\mathbf{l}_{2}$ &             $\mathbf{l}_{3}$ & $\mathllap{-}\mathbf{p}$ &             $\bm{\pi}_{1}$ & $\bm{\pi}_{2}$ & $\mathllap{-}\bm{\pi}_{3}$\\ 
\end{tabular}
\end{ruledtabular}
\end{table}

\textcolor{newtext}{Next, we will use an effective and elegant symmetry approach for physical phenomena in collinear antiferromagnets developed by Turov~\cite{turov2001symmetry} and successfully utilized in many papers, e.g. in Refs.~\cite{kimel2024optical,mostovoy2024multiferroics}.
This approach enables the use of the antiferromagnetic vector $\mathbf{l}$, which, despite being axial, sometimes behaves like a polar, within a symmetry-based analysis.}
From a symmetry point of view, the trigonal space group of \CrO{} can be represented by three generators of the group (independent symmetry elements), which include an inversion center $\overline{1}$ ($x \rightarrow -x$, $y \rightarrow -y$, $z \rightarrow -z$), a three-fold axis $3_{z} \parallel z$ ($x \rightarrow -\frac{1}{2} \, [x - \sqrt{3} \, y]$, $y \rightarrow -\frac{1}{2} \, [\sqrt{3} \, x + y]$, $z \rightarrow z$) and a two-fold axis $2_{x} \parallel x$ ($x \rightarrow x$, $y \rightarrow -y$, $z \rightarrow -z$)~\cite{turov2001symmetry}.
The action of these symmetry elements can be illustrated, if we keep in mind that \textcolor{newtext2}{in the case of four magnetic ions in a unit cell of \CrO{} we should define four magnetic and four electric basis vectors} by the following way:
\begin{equation}
\label{eq:m_l_vectors}
\begin{gathered}
\mathbf{m} = \frac{\mathbf{m}_{1} + \mathbf{m}_{2} + \mathbf{m}_{3} + \mathbf{m}_{4}}{4 \, m_{0}},\\
\mathbf{l}_{1} = \frac{\mathbf{m}_{1} - \mathbf{m}_{2} - \mathbf{m}_{3} + \mathbf{m}_{4}}{4 \, m_{0}},\\
\mathbf{l}_{2} = \frac{\mathbf{m}_{1} - \mathbf{m}_{2} + \mathbf{m}_{3} - \mathbf{m}_{4}}{4 \, m_{0}},\\
\mathbf{l}_{3} = \frac{\mathbf{m}_{1} + \mathbf{m}_{2} - \mathbf{m}_{3} - \mathbf{m}_{4}}{4 \, m_{0}},
\end{gathered}
\end{equation}
and
\begin{equation}
\label{eq:antiferroelectric vectors}
\begin{gathered}
\mathbf{p} = \frac{\mathbf{d}_{1} + \mathbf{d}_{2} + \mathbf{d}_{3} + \mathbf{d}_{4}}{4 \, d_{0}},\\
\bm{\pi}_{1} = \frac{\mathbf{d}_{1} - \mathbf{d}_{2} - \mathbf{d}_{3} + \mathbf{d}_{4}}{4 \, d_{0}},\\
\bm{\pi}_{2} = \frac{\mathbf{d}_{1} - \mathbf{d}_{2} + \mathbf{d}_{3} - \mathbf{d}_{4}}{4 \, d_{0}},\\
\bm{\pi}_{3} = \frac{\mathbf{d}_{1} + \mathbf{d}_{2} - \mathbf{d}_{3} - \mathbf{d}_{4}}{4 \, d_{0}}.
\end{gathered}
\end{equation}
The group generators perform permutations of magnetic ions in the unit cell, which in turn lead to the transformation of magnetic [Eq.~\eqref{eq:m_l_vectors}] and electric [Eq.~\eqref{eq:antiferroelectric vectors}] basis vectors, as listed in Table~\ref{tab:generators}.

The generators of the group can be divided into two types based on their permutation properties with respect to the selected position of the magnetic ions.
A symmetry element is labelled with an index $(+)$ if it permutes ions to positions belonging to the same magnetic sublattice with equally oriented magnetizations, and with $(-)$ if the resulting sublattice is opposite.
According to these rules, we can write the symmetry elements adopted as the generators of the space groups $\overline{1}(-) \, 3_{z}(+) \, 2_{x}(-)$ for \CrO, as can be seen in Fig.~\ref{fig:structure}.
\textcolor{newtext2}{From Table~\ref{tab:generators}, we see that $3_{z}(+)$ transforms $\mathbf{m}$ and $\mathbf{l}_{2}$ in the same way, whereas under the action of $\overline{1}(-)$ and $2_{x}(-)$ the vector $\mathbf{l}_{2}$ changes sign.
It is worth noting that in a similar way these indices relate polarization $\mathbf{p}$ and antiferroelectric vector $\bm{\pi}$.} 
\textcolor{newtext}{Thus, the $(+)$ and $(-)$ indices are relevant only for antiferromagnetic $\mathbf{l} = \mathbf{l}_{2}$ and antiferroelectric $\bm{\pi} = \bm{\pi}_{2}$ vectors, which, under the action of a symmetry element, are transformed like axial and polar vectors multiplied by the sign of the index, respectively.}
For example, $\overline{1}(-) \, \mathbf{m} = \mathbf{m}$ and $\overline{1}(-) \, \mathbf{p} = -\mathbf{p}$, whereas $\overline{1}(-) \; \mathbf{l} = - \mathbf{l}$ and $\overline{1}(-) \, \bm{\pi} = \bm{\pi}$.
Note that in our case the magnetic sublattices coincide with electric dipole sublattices (see Fig.~\ref{fig:structure}), but if these sublattices do not match then the indices can be different, e.g. $\overline{1}(-)$ for magnetic and $\overline{1}(+)$ for electric basis vectors.
Thus, the set of generators, written in this way, and called an exchange magnetic structure, provides all the necessary information to find the invariant form of a thermodynamic potential or a material tensor in terms of magnetization $\mathbf{m}$, antiferromagnetic vector $\mathbf{l}$, polarization $\mathbf{p}$, antiferroelectric vector $\bm{\pi}$, electric current $\mathbf{j}$, etc., for antiferromagnets as detailed in Refs.~\cite{turov2001symmetry,turov2005new}. 

\begin{table}
\caption{\label{tab:symmetry} Transformation of the magnetization $\mathbf{m}$, antiferromagnetic vector $\mathbf{l}$, electric current $\mathbf{j}$, polarization $\mathbf{p}$ and antiferroelectric vector $\bm{\pi}$ under the action of symmetry elements $\overline{1}(-)$, $3_{z}(+)$, $2_{x}(-)$.}
\begin{ruledtabular}
\begin{tabular}{cccc}
\makecell{Dynamical\\ variable} &   $\overline{1}(-)$ &                                        $3_{z}(+)$ &            $2_{x}(-)$ \\ \midrule
                        $m_{x}$ &             $m_{x}$ &     $\mathllap{-}(m_{x} - \sqrt{3} \, m_{y}) / 2$ &               $m_{x}$ \\
                        $m_{y}$ &             $m_{y}$ &     $\mathllap{-}(\sqrt{3} \, m_{x} + m_{y}) / 2$ &   $\mathllap{-}m_{y}$ \\
                        $m_{z}$ &             $m_{z}$ &                                           $m_{z}$ &   $\mathllap{-}m_{z}$ \\ \midrule
                        $l_{x}$ & $\mathllap{-}l_{x}$ &     $\mathllap{-}(l_{x} - \sqrt{3} \, l_{y}) / 2$ &   $\mathllap{-}l_{x}$ \\
                        $l_{y}$ & $\mathllap{-}l_{y}$ &     $\mathllap{-}(\sqrt{3} \, l_{x} + l_{y}) / 2$ &               $l_{y}$ \\
                        $l_{z}$ & $\mathllap{-}l_{z}$ &                                           $l_{z}$ &               $l_{z}$ \\ \midrule
                        $j_{x}$ & $\mathllap{-}j_{x}$ &     $\mathllap{-}(j_{x} - \sqrt{3} \, j_{y}) / 2$ &               $j_{x}$ \\
                        $j_{y}$ & $\mathllap{-}j_{y}$ &     $\mathllap{-}(\sqrt{3} \, j_{x} + j_{y}) / 2$ &   $\mathllap{-}j_{y}$ \\
                        $j_{z}$ & $\mathllap{-}j_{z}$ &                                           $j_{z}$ &   $\mathllap{-}j_{z}$ \\ \midrule
                        $p_{x}$ & $\mathllap{-}p_{x}$ &     $\mathllap{-}(p_{x} - \sqrt{3} \, p_{y}) / 2$ &               $p_{x}$ \\
                        $p_{y}$ & $\mathllap{-}p_{y}$ &     $\mathllap{-}(\sqrt{3} \, p_{x} + p_{y}) / 2$ &   $\mathllap{-}p_{y}$ \\
                        $p_{z}$ & $\mathllap{-}p_{z}$ &                                           $p_{z}$ &   $\mathllap{-}p_{z}$ \\ \midrule
                      $\pi_{x}$ &           $\pi_{x}$ & $\mathllap{-}(\pi_{x} - \sqrt{3} \, \pi_{y}) / 2$ & $\mathllap{-}\pi_{x}$ \\ 
                      $\pi_{y}$ &           $\pi_{y}$ & $\mathllap{-}(\sqrt{3} \, \pi_{x} + \pi_{y}) / 2$ &             $\pi_{y}$ \\
                      $\pi_{z}$ &           $\pi_{z}$ &                                    $\pi_{z}$      &             $\pi_{z}$ \\ 
\end{tabular}
\end{ruledtabular}
\end{table}

Analysis of the results of the action of symmetry elements for $\overline{1}(-) \, 3_{z}(+) \, 2_{x}(-)$ in \CrO{} (see Table~\ref{tab:symmetry}) reveals that the combination of components of the antiferromagnetic vector and electric current $l_{x} \, j_{y} - l_{y} \, j_{x}$ remains invariant, i.e., \textcolor{newtext2}{it is transformed according to the fully symmetric $\Gamma_{1}$ irreducible representation}.
Notably, this expression represents the $z$ component of the cross product $[\mathbf{l} \times \mathbf{j}]_{z} = l_{x} \, j_{y} - l_{y} \, j_{x}$.  
To form a scalar for a free energy, this cross product must be dotted with a vector \textcolor{newtext2}{of $\Gamma_{1}$ symmetry} directed along the $z$ axis, forming a triple product, \textcolor{newtext2}{and besides, invariance with respect to time reversal must be fulfilled}.
\textcolor{newtext2}{According to symmetry, only $t$-even antiferroelectric polar vector $\bm{\pi} \parallel z$ that does not change sign under space inversion $\overline{1}$ satisfies such requirements (see Table~\ref{tab:symmetry}).}   
As a result, the free energy related to the interaction of electric current $\mathbf{j}$ with the antiferromagnetic vector $\mathbf{l}$ in \CrO{}, which is related to the spin-orbit torque (SOT), can be written as follows 
\begin{equation}
\label{eq:SOT}
\mathcal{F}_{\mathrm{SOT}} \simeq \bm{\pi} \cdot [\mathbf{l} \times \mathbf{j}].
\end{equation}

The symmetry requirements for the existence of the linear magnetoelectric effect and SOT are substantially similar. 
For the linear magnetoelectric effect to be symmetry-allowed in a collinear antiferromagnet with inversion, the space inversion operation must be odd $\overline{1}(-)$, meaning it connects two opposite magnetic sublattices as depicted in Fig.~\ref{fig:structure}(a); this results in the antiferromagnetic vector $\mathbf{l}$ being an axial vector that changes sign under space inversion $\overline{1}$.
For SOT, in our view, the essential requirement is the presence of a nonzero antiferroelectric vector $\bm{\pi}$ in a collinear antiferromagnet possessing space inversion $\overline{1}$ [see Fig.~\ref{fig:structure}(c)].
\textcolor{newtext2}{It is worth noting that, for the electric dipole order vector $\bm{\pi}$ the inversion can be only odd $\overline{1}(-)$, because electric dipoles $\mathbf{d}$ are determined by the positions of ions and they must be equidistant from the inversion center.
In contrast, for the antiferromagnetic vector $\mathbf{l}$, the inversion can be odd $\overline{1}(-)$ or even $\overline{1}(+)$, since it depends on the spin ordering.} 
We note that this $\bm{\pi}$ vector in Eq.~\eqref{eq:SOT} acts as the source of the staggered Rashba electric field $\mathbf{E}_{\mathrm{R}}$, which is key for SOT~\cite{bihlmayer2022rashba}.

In order for SOT to be symmetry resolved, the cross product $\bm{\pi} \cdot [\mathbf{l} \times \mathbf{j}]$ must contain invariant combinations.
This necessitates that the space inversion links opposite or co-directional magnetic and electric dipole sublattices, i.e., must be odd $\overline{1}(-)$ simultaneously for $\mathbf{l}$ and for $\bm{\pi}$, \textcolor{newtext2}{meaning that it reverses the axial vector $\mathbf{l}$ but leaves the polar vector $\bm{\pi}$ unchanged}.
As already stated, in \CrO{} with its $\overline{1}(-)$ symmetry, the magnetic and electric dipole sublattices are co-directional as shown in Figs.~\ref{fig:structure}(a) and~\ref{fig:structure}(c) allowing both the linear magnetoelectric effect and SOT.
In contrast, the SOT driven magnetic dynamics \textcolor{newtext}{is} symmetry forbidden in the isostructural antiferromagnet hematite $\alpha$-$\mathrm{Fe}_{2}\mathrm{O}_{3}$ with different spin ordering and the exchange magnetic structure $\overline{1}(+) \, 3_{z}(+) \, 2_{x}(-)$ but with the same electric dipole structure as in \CrO{} [i.e., $\overline{1}(+)$ for $\mathbf{l}$, but $\overline{1}(-)$ for $\bm{\pi}$] which eliminates the invariant $\pi_{z} \, (l_{x} \, j_{y} - l_{y} \, j_{x})$.
The linear magnetoelectric effect is also forbidden in $\alpha$-$\mathrm{Fe}_{2}\mathrm{O}_{3}$ as a result of its $\overline{1}(+)$ symmetry for $\mathbf{l}$.
\textcolor{newtext2}{It should be noted that, the detailed symmetry analysis of SOT in metallic antiferromagnets is provided in Ref.~\cite{zelezny2014relativistic}.}

\begin{figure}
\centering
\includegraphics[width=1\columnwidth]{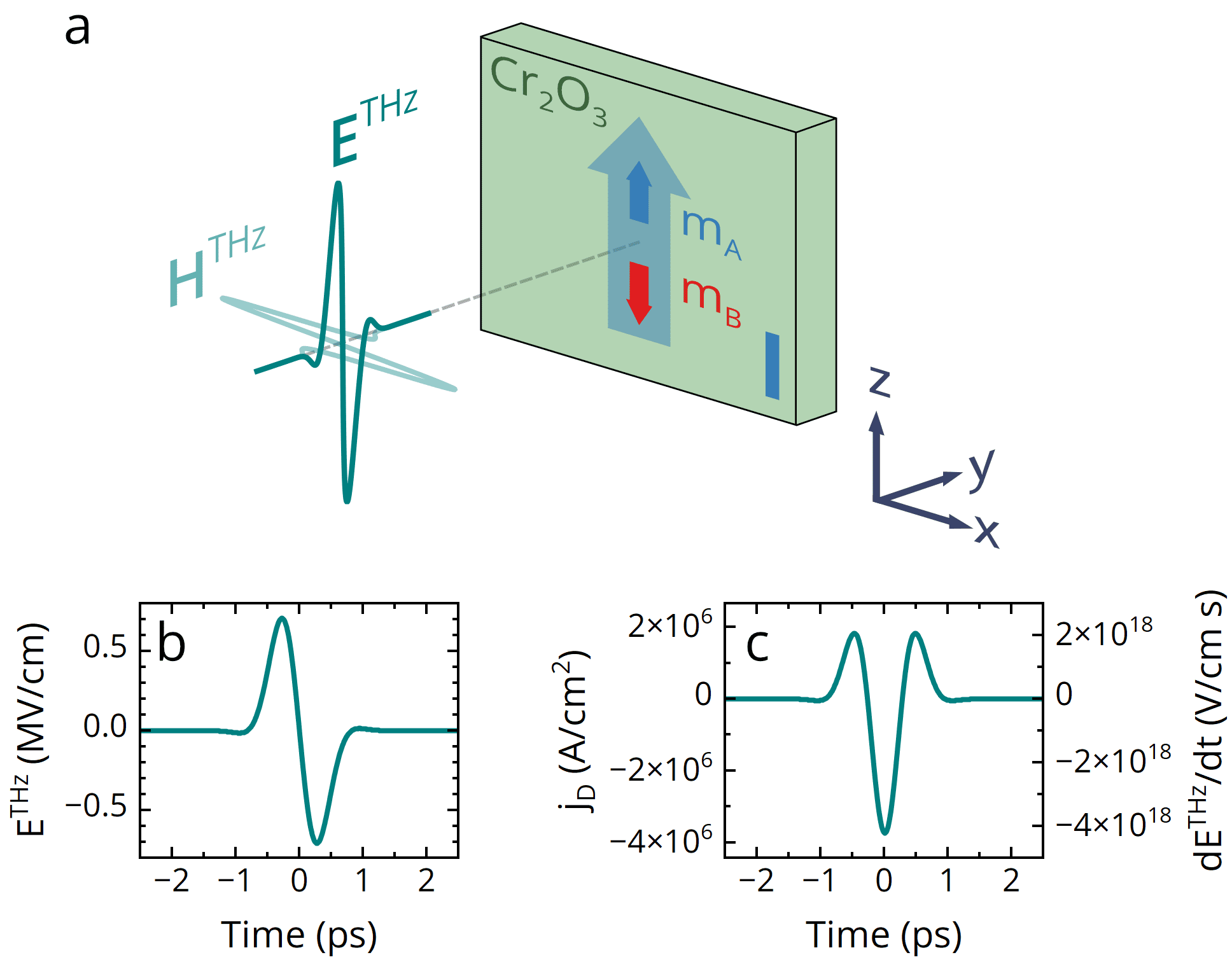}
\caption{\label{fig:displacement_current}
(a)~Geometry of the considered THz experiment, in which dynamics of the antiferromagnetic vector $\mathbf{l} = \mathbf{m}_{\mathrm{A}} - \mathbf{m}_{\mathrm{B}}$ in \CrO{} is driven by a THz nearly single cycle pulse.
Time traces of (b)~the THz electric field $\mathbf{E}^{\mathrm{THz}}$ and (c)~the resulting displacement current $\mathbf{j}_{\mathrm{D}} \propto \dot{\mathbf{E}}$ along with the time derivative $\dot{\mathbf{E}}$.
}
\end{figure}

It is worth noting that the coupling between the electric current $\mathbf{j}$ and antiferromagnetic vector $\mathbf{l}$ is well known in metallic antiferromagnets such as $\mathrm{Mn}_{2}\mathrm{Au}$ and $\mathrm{CuMnAs}$ and it is related to N{\'e}el spin-orbit torque~\cite{wadley2016electrical,bodnar2018writing,manchon2019current,troncoso2019antiferromagnetic,selzer2022current,behovits2023terahertz,kaspar2021quenching,freimuth2021laser,ross2024ultrafast,olejnik2024quench}.
It is clear that, since \CrO{} single crystal is a good insulator, the electric currents flowing through it are negligible.   
However, according to Maxwell's equations, there is a displacement current in insulators that is caused by the motion of bound charges and, according to Maxwell's equations, like electric current, results in a magnetic field~\cite{landau1984electrodynamics,sivukhin1996course,jefimenko1966electricity}.
\textcolor{newtext}{We note that the displacement current $\mathbf{j}_{\mathrm{D}}$ holds significant potential in spintronics, for example, for inducing magnetization dynamics in magnetic tunnel junctions~\cite{safeer2024magnetization} and ultrasound in magnetic insulators~\cite{buchel1997electromagnetic}.}
The displacement current in the Gaussian system of units has the form $\mathbf{j}_{\mathrm{D}} = \cfrac{\varepsilon}{4 \, \pi} \, \dot{\mathbf{E}}$, where $\varepsilon$ is the dielectric permittivity. 
We anticipate that the dielectric permittivity in the relevant frequency range will exhibit negligible deviation from the static dielectric permittivity. The latter is described by a diagonal tensor with two independent components, $\varepsilon_{x} = 10.33$ and $\varepsilon_{z} = 11.93$~\cite{lucovsky1977infrared}, the difference between which we disregard for simplicity.
We consider an experiment with a near single-cycle THz pulse with a duration of about 2\,ps, a spectral maximum at 0.6\,THz, and a peak electric field strength of 760\,kV/cm, as shown in Figs.~\ref{fig:displacement_current}(a) and~\ref{fig:displacement_current}(b).
The time derivative reveals the displacement current $\mathbf{j} \propto \dot{\mathbf{E}}^{\mathrm{THz}}$ induced in \CrO{} by this THz pulse, with a peak amplitude of about $4 \times 10^{6}$\,A/cm$^{2}$, as shown in Fig.~\ref{fig:displacement_current}(c).
Thus, it is interesting to explore how the displacement current $\mathbf{j}_{\mathrm{D}}$ induced by the THz electric field $\mathbf{E}^{\mathrm{THz}}$ drives the dynamics of the antiferromagnetic vector $\mathbf{l}$ through spin-orbit torque in the magnetoelectric \CrO.

\subsection{Spin Dynamics}

To reveal the key similarities and differences between the linear magnetoelectric effect and SOT, we develop a model of spin dynamics driven by the THz pulses in \CrO{} with the antiferromagnetic vector $\mathbf{l} \parallel z$ using a non-standard spherical coordinate system where the polar angle $\vartheta$ counts from the $y$ axis and the azimuthal angle $\varphi$ lies in the $xz$ plane and counts from the $x$ axis.
The sublattice magnetizations in this case have the form $\mathbf{m_{\mathrm{A,B}}} = (\sin{\vartheta_{\mathrm{A,B}}} \, \cos{\varphi_{\mathrm{A,B}}}, \cos{\vartheta_{\mathrm{A,B}}},  \sin{\vartheta_{\mathrm{A,B}}} \, \sin{\varphi_{\mathrm{A,B}}})$.
Then, the spherical angles of $\mathbf{m_{\mathrm{A}}}$ and $\mathbf{m_{\mathrm{B}}}$ are parametrized as follows  
\begin{equation}
\label{eq:canted_angles}
\begin{gathered}
    \vartheta_{\mathrm{A}} = \vartheta - \epsilon, \quad \vartheta_{\mathrm{B}} = \pi - \vartheta - \epsilon,\\
    \varphi_{\mathrm{A}} = \varphi + \beta, \quad \varphi_{\mathrm{B}} = \pi + \varphi - \beta,
\end{gathered}
\end{equation}
where small canting angles $\epsilon \ll 1$ and $\beta \ll 1$ are introduced. 
This allowed us to expand the net magnetization $\mathbf{m}$ and antiferromagnetic $\mathbf{l}$ vector Cartesian components in series with respect to $\varepsilon$ and $\beta$ angles
\begin{equation}
\label{eq:vectors}
\begin{gathered}
    m_{x} \approx  - \beta \, \sin{\vartheta} \, \sin{\varphi} - \epsilon \, \cos{\vartheta} \, \cos{\varphi},\\
    m_{y} \approx \epsilon \, \sin{\vartheta},\\
    m_{z} \approx \beta \, \sin{\vartheta} \, \cos{\varphi} - \epsilon \, \cos{\vartheta} \, \sin{\varphi},\\
    l_{x} \approx \sin{\vartheta} \, \cos{\varphi},\\
    l_{y} \approx \cos{\vartheta},\\
    l_{z} \approx \sin{\vartheta} \, \sin{\varphi}.\\
\end{gathered}
\end{equation}

The ground state of the antiferromagnetic vector $\mathbf{l} \parallel z$ is defined by the angles $\vartheta_{0} = \cfrac{\pi}{2}$ and $\varphi_{0} = \pm\cfrac{\pi}{2}$, where $\pm$ denotes two different antiferromagnetic domains.
Near the ground state, the angles can be expressed as $\vartheta = \vartheta_{0} + \vartheta_{1}$ and $\varphi = \varphi_{0} + \varphi_{1}$, where $\vartheta_{1} \ll 1$ and $\vartheta_{1} \ll 1$.
Then the following expansions are relevant
\begin{equation}
\begin{gathered}
    \sin{\vartheta} \approx 1 - \frac{\vartheta_{1}^2}{2}, \quad \cos{\vartheta}\approx - \vartheta_{1}, \\
    \sin{\varphi} \approx \pm \Bigl(1 - \frac{\varphi_{1}^2}{2} \Bigr), \quad \cos{\varphi} \approx \mp\varphi_{1}.
\end{gathered}
\end{equation}
This allows us to represent Cartesian components of the $\mathbf{m}$ and $\mathbf{l}$ vectors from Eq.~\eqref{eq:vectors} in the form
\begin{equation}
\label{eq:small_vectors}
\begin{gathered}
    m_{x} \approx  \mp \beta,\\
    m_{y} \approx \epsilon,\\
    m_{z} \approx \pm (\epsilon \, \vartheta_{1} - \beta \, \varphi_{1}),\\
    l_{x} \approx \mp \varphi_{1},\\
    l_{y} \approx - \vartheta_{1},\\
    l_{z} \approx \pm \Bigl(1 - \cfrac{\varphi_{1}^{2}}{2} - \cfrac{\vartheta_{1}^{2}}{2} \Bigr).\\
\end{gathered}
\end{equation}

First, we need to define expressions for the energy of the spin system of \CrO{} near the ground state using Eq.~\eqref{eq:small_vectors}.
The kinetic energy of a double-sublattice antiferromagnet \textcolor{newtext2}{determined through the Berry phase gauge $\gamma_{\mathrm{Berry}} = (1 - \cos{\vartheta_{\mathrm{A}}}) \, \dot{\varphi}_{\mathrm{A}} + (1 - \cos{\vartheta_{\mathrm{B}}}) \, \dot{\varphi}_{\mathrm{B}}$} in the first order in $\epsilon$ and $\beta$ can be represented as~\cite{fradkin2013field,zvezdin2024giant}
\begin{equation}
\label{eq:T}
    T = \frac{m_{0}}{\gamma} \, (\epsilon \dot{\varphi}_{1} + \beta \dot{\vartheta}_{1}),
\end{equation}
where $\gamma$ is the gyromagnetic ratio.
\textcolor{newtext2}{This expression for small deviations of $\mathbf{l}$ from its equilibrium position is equivalent to the well-known expression for the kinetic energy of an antiferromagnet $T \propto \dot{\mathbf{l}}^{2}$~\cite{andreev1980symmetry,zvezdin2024giant}.}

The exchange energy taking into account Eq.~\eqref{eq:canted_angles} has the next form in the second order of $\epsilon$ and $\beta$ up to a constant term
\begin{equation}
\label{eq:J}
    U_{\mathrm{Ex}} = \lambda_{\mathrm{Ex}} \, m_{0}^{2} \, \mathbf{m}_{\mathrm{A}} \cdot \mathbf{m}_{\mathrm{B}} \approx 2 \, \lambda_{\mathrm{Ex}} \, m_{0}^{2} \, (\epsilon^{2} + \beta^{2}),
\end{equation}
where $\lambda_{\mathrm{Ex}} = 1 / \chi_{\perp}$ is the exchange interaction between neighbouring spins of $\mathrm{Cr}^{3+}$ ions and $\chi_{\perp}$ is the perpendicular magnetic susceptibility.
The energy of the uniaxial magnetic anisotropy in the second order in $\varphi_{1}$ and $\vartheta_{1}$ up to a constant term is 
\begin{equation}
\label{eq:A}
\begin{gathered}
    U_{\mathrm{A}} = - K \, l_{z}^{2} \approx K \, (\varphi_{1}^{2} + \vartheta_{1}^{2}),
\end{gathered}
\end{equation}
where $K$ is the uniaxial anisotropy constant.

The Zeeman interaction of the antiferromagnet with the external magnetic field $\mathbf{H}$, taking into account Eq.~\eqref{eq:small_vectors}, has the following form  
\begin{equation}
\label{eq:Zeeman}
\begin{gathered}
    U_{\mathrm{Z}} = - m_{0} \, \mathbf{m} \cdot \mathbf{H} = - m_{0} \, \left[ m_{x} H_{x} + m_{y} H_{y} + m_{z} H_{z}\right] \\
    \approx - m_{0} \, \left[\mp \beta \, H_{x} + \epsilon \, H_{y} \pm (\epsilon \, \vartheta_{1} - \beta \, \varphi_{1}) \, H_{z} \right].
\end{gathered}
\end{equation}

According to Ref.~\cite{turov2001symmetry}, the general expression for the linear magnetoelectric interaction energy in \CrO{} in the second order in small angles is   
\begin{equation}
\label{eq:ME}
\begin{gathered}
    U_{\mathrm{ME}} = - \lambda_{\mathrm{ME}1} \, m_{0} \, [(l_{x} \, m_{y} + l_{y} \, m_{x}) \, E_{x} + (l_{x} \, m_{x} - l_{y} \, m_{y}) \, E_{y}] \\
    - \lambda_{\mathrm{ME}2} \, m_{0} \, m_{z} \, (l_{x} \, E_{x} + l_{y} \, E_{y})  - \lambda_{\mathrm{ME}3} \, m_{0} \, l_{z} \, (m_{x} \, E_{x} + m_{y} \, E_{y}) \\
    - \lambda_{\mathrm{ME}4} \, m_{0} \, (l_{x} \, m_{x} + l_{y} \, m_{y}) \, E_{z} - \lambda_{\mathrm{ME}5} \, m_{0} \, l_{z} \, m_{z} \, E_{z} \\ \approx 
    - \lambda_{\mathrm{ME}1} \, m_{0} \, [\pm(\beta \, \vartheta_{1} - \epsilon \, \vartheta_{1}) \, E_{x} + (\beta \, \varphi_{1} + \epsilon \, \varphi_{1}) \, E_{y}]\\
    - \lambda_{\mathrm{ME}3} \, m_{0} \, (- \beta \, E_{x} \pm \epsilon \, E_{y}) - \lambda_{\mathrm{ME}4} \, m_{0} \, (\beta \, \varphi_{1} - \epsilon \, \vartheta_{1}) \, E_{z}\\
    - \lambda_{\mathrm{ME}5} \, m_{0} \, (\epsilon \, \vartheta_{1} - \beta \, \varphi_{1}) \, E_{z},
\end{gathered}
\end{equation}
where $\lambda_{\mathrm{ME}1-5}$ are magnetoelectric parameters.
Strictly speaking, the electric polarization $\mathbf{P}$ should be written instead of electric field $\mathbf{E}$ in this expression, since $\mathbf{P}$ is a dynamical variable of the medium. 
But for insulating crystals, the polarization response to an electric field is given by the linear relation $\mathbf{P} = \cfrac{\varepsilon - 1}{4 \, \pi} \, \mathbf{E}$.

Finally, the SOT interaction energy [Eq.~\eqref{eq:SOT}] can be expressed as
\begin{equation}
\label{eq:U_SOT}
\begin{gathered}
    U_{\mathrm{SOT}} = - \lambda_{\mathrm{SOT}} \, m_{0} \, \pi_{z} \, (l_{x} \, j_{y} - l_{y} \, j_{x})\\
    \approx - \lambda_{\mathrm{SOT}} \, m_{0} \, \pi_{z} \, \varepsilon \, / \, (4 \, \pi) \, (\mp\varphi_{1} \, \dot{E}_{y} + \vartheta_{1} \, \dot{E}_{x}),
\end{gathered}
\end{equation}
where $\lambda_{\mathrm{SOT}}$ is the SOT parameter.

To describe the spin dynamics in \CrO{} we employ a Lagrangian using Eqs.~\eqref{eq:T}, \eqref{eq:J}, \eqref{eq:A}, \eqref{eq:Zeeman}, \eqref{eq:U_SOT} and \eqref{eq:ME}
\begin{equation}
\begin{gathered}
\label{eq:L}
    \mathcal{L} = T - U_{\mathrm{Ex}} - U_{\mathrm{A}} - U_{\mathrm{Z}} - U_{\mathrm{ME}} - U_{\mathrm{SOT}} \\
    = \frac{m_{0}}{\gamma} \, (\epsilon \dot{\varphi}_{1} + \beta \dot{\vartheta}_{1}) - 2 \, \lambda_{\mathrm{Ex}} \, m_{0}^{2} \, (\epsilon^{2} + \beta^{2}) - K \, (\varphi_{1}^{2} + \vartheta_{1}^{2}) \\
    + m_{0} \, \left[\mp \beta \, H_{x} + \epsilon \, H_{y} \pm (\epsilon \, \vartheta_{1} - \beta \, \varphi_{1}) \, H_{z} \right] \\
    + \lambda_{\mathrm{ME}1} \, m_{0} \, [\pm(\beta \, \vartheta_{1} - \epsilon \, \varphi_{1}) \, E_{x} + (\beta \, \varphi_{1} + \epsilon \, \vartheta_{1}) \, E_{y}] \\
    + \lambda_{\mathrm{ME}3} \, m_{0} \, (- \beta \, E_{x} \pm \epsilon \, E_{y}) + \lambda_{\mathrm{ME}4} \, m_{0} \, (\beta \, \varphi_{1} - \epsilon \, \vartheta_{1}) \, E_{z} \\
    + \lambda_{\mathrm{ME}5} \, m_{0} \, (\epsilon \, \vartheta_{1} - \beta \, \varphi_{1}) \, E_{z}\\
    + \lambda_{\mathrm{SOT}} \, m_{0} \, \pi_{z} \, \varepsilon \, / \, (4 \, \pi) \, (\mp\varphi_{1} \, \dot{E}_{y} + \vartheta_{1} \, \dot{E}_{x}).
\end{gathered}
\end{equation}
Then using the Lagrangian~\eqref{eq:L}, we solve the Euler-Lagrange equations
\begin{equation}
    \frac{d}{dt} \frac{\partial \mathcal L}{\partial \dot q_{i}} - \frac{\partial \mathcal L}{\partial q_{i}} = 0,
\end{equation}
where $q_{i}$ for $i=1$--$4$ are order parameters $\epsilon$, $\varphi_{1}$, $\beta$, and $\vartheta_{1}$, respectively.
\textcolor{newtext}{Note that the Gilbert damping is omitted from the Euler-Lagrange equations due to its negligible impact on our results.
This is because the relevant torques are field-like, not dissipative-like, and the magnon damping time in \CrO{} ($\tau_{\mathrm{M}} \approx 700$\,ps~\cite{mukhin1997bwo}) vastly exceeds the  time delay ($\Delta \tau \approx100$\,ps~\cite{bilyk2025control}) in typical THz pump-probe experiments.}   

As a result, we obtain a system of four coupled differential equations describing the spin dynamics in the magnetoelectric \CrO{} 
\begin{equation}
\begin{gathered}
\label{eq:epsilon_full}
    \dot{\epsilon} + \omega_{\mathrm{A}} \, \varphi_{1} \pm \gamma \, \beta \, H_{z} \pm \Tilde{\lambda}_{\mathrm{ME}1} \, \epsilon \, E_{x} - \Tilde{\lambda}_{\mathrm{ME}1} \beta \, E_{y} \\
    - (\Tilde{\lambda}_{\mathrm{ME}4} - \Tilde{\lambda}_{\mathrm{ME}5}) \, \beta \, E_{z} = \mp\Tilde{\lambda}_{\mathrm{SOT}} \, \dot{E}_{y},
\end{gathered}    
\end{equation}
\begin{equation}
\begin{gathered}
\label{eq:phi1_full}
    \dot{\varphi}_{1} - \omega_{\mathrm{Ex}} \, \epsilon \pm \gamma \, \vartheta_{1} \, H_{z} \mp \Tilde{\lambda}_{\mathrm{ME}1} \, \varphi_{1} \, E_{x} \\
    + \Tilde{\lambda}_{\mathrm{ME}1} \, \vartheta_{1} \, E_{y} - (\Tilde{\lambda}_{\mathrm{ME}4} - \Tilde{\lambda}_{\mathrm{ME}5}) \, \vartheta_{1} \, E_{z} = \mp \Tilde{\lambda}_{\mathrm{ME}3} \, E_{y} - \gamma \, H_{y},
\end{gathered}    
\end{equation}
\begin{equation}
\begin{gathered}
\label{eq:beta_full}
    \dot{\beta} + \omega_{\mathrm{A}} \, \vartheta_{1} \mp \gamma \, \epsilon \, H_{z} \mp \Tilde{\lambda}_{\mathrm{ME}1} \, \beta \, E_{x} - \Tilde{\lambda}_{\mathrm{ME}1} \, \epsilon \, E_{y} \\
    + (\Tilde{\lambda}_{\mathrm{ME}4} - \Tilde{\lambda}_{\mathrm{ME}5}) \, \epsilon \, E_{z} = \Tilde{\lambda}_{\mathrm{SOT}} \, \dot{E}_{x},
\end{gathered}    
\end{equation}
\begin{equation}
\begin{gathered}
\label{eq:theta_1_full}
    \dot{\vartheta}_{1} - \omega_{\mathrm{Ex}} \, \beta \mp \gamma \, \varphi_{1} \, H_{z} \pm \Tilde{\lambda}_{\mathrm{ME}1} \, \vartheta_{1} \, E_{x} + \Tilde{\lambda}_{\mathrm{ME}1} \, \varphi_{1} \, E_{y}\\
    + (\Tilde{\lambda}_{\mathrm{ME}4} - \Tilde{\lambda}_{\mathrm{ME}5}) \, \varphi_{1} \, E_{z} = \Tilde{\lambda}_{\mathrm{ME}3} \, E_{x} \pm \gamma \, H_{x},
\end{gathered}    
\end{equation}
where 
$\omega_{\mathrm{A}} = \gamma \, H_{\mathrm{A}} = 2 \, \gamma \, K / m_{0} \simeq 1.2 \times 10^{10}$\,rad/s~\cite{foner1963high},
$\omega_{\mathrm{Ex}} = \gamma \, H_{\mathrm{Ex}} = 4 \, \gamma \, \lambda_{\mathrm{Ex}} \, m_{0} \simeq 8.6 \times 10^{13}$\,rad/s~\cite{foner1963high}, 
$\Tilde{\lambda}_{\mathrm{ME}1-5} = \gamma \, \lambda_{\mathrm{ME}1-5}$,
$\Tilde{\lambda}_{\mathrm{SOT}} = \gamma \, \lambda_{\mathrm{SOT}} \, \pi_{z} \, \varepsilon \, / \, (4 \, \pi)$.
The system of differential Eqs.~\eqref{eq:epsilon_full}--\eqref{eq:theta_1_full} describes the dynamics of a doubly degenerate magnon in \CrO{} characterized by the frequency $\omega_{\mathrm{M}} = \sqrt{\omega_{\mathrm{Ex}} \, \omega_{\mathrm{A}}}$, driven by the $\mathbf{E}(t)$ and $\mathbf{H}(t)$ torques appearing on the right-hand side of the equations.     
Besides, there are intrinsic parametric $\mathbf{E}(t)$ and $\mathbf{H}(t)$ torques on the left side of these equations, which we neglect for simplicity since they do not significantly affect the effects at the field values of interest. 
Thus, the system of Eqs.~\eqref{eq:epsilon_full}--\eqref{eq:theta_1_full} takes a more straightforward form
\begin{equation}
\begin{gathered}
\label{eq:epsilon_short}
    \dot{\epsilon} + \omega_{\mathrm{A}} \, \varphi_{1} = \mp \Tilde{\lambda}_{\mathrm{SOT}} \, \dot{E}_{y},
\end{gathered}    
\end{equation}
\begin{equation}
\begin{gathered}
\label{eq:phi1_short}
    \dot{\varphi}_{1} - \omega_{\mathrm{Ex}} \, \epsilon = \mp \Tilde{\lambda}_{\mathrm{ME}3} \, E_{y} - \gamma \, H_{y},
\end{gathered}    
\end{equation}
\begin{equation}
\begin{gathered}
\label{eq:beta_short}
    \dot{\beta} + \omega_{\mathrm{A}} \, \vartheta_{1} = \Tilde{\lambda}_{\mathrm{SOT}} \, \dot{E}_{x},
\end{gathered}    
\end{equation}
\begin{equation}
\begin{gathered}
\label{eq:theta_1_short}
    \dot{\vartheta}_{1} - \omega_{\mathrm{Ex}} \, \beta = \Tilde{\lambda}_{\mathrm{ME}3} \, E_{x} \pm \gamma \, H_{x},
\end{gathered}    
\end{equation}
which can be reduced to four second order differential equations 
describing the dynamics of the magnetization $\mathbf{m}$ and antiferromagnetic vector $\mathbf{l}$ components 
\begin{equation}
\begin{gathered}
\label{eq:epsilon_SODE}
    \Ddot{\epsilon} + \omega_{\mathrm{A}} \, \omega_{\mathrm{Ex}} \epsilon = \mp \Tilde{\lambda}_{\mathrm{SOT}} \, \Ddot{E}_{y} \pm \omega_{\mathrm{A}} \, \Tilde{\lambda}_{\mathrm{ME}3} \, E_{y}  + \gamma \, \omega_{\mathrm{A}} \, H_{y},
\end{gathered}    
\end{equation}
\begin{equation}
\begin{gathered}
\label{eq:phi1_SODE}
    \Ddot{\varphi}_{1} + \omega_{\mathrm{A}} \, \omega_{\mathrm{Ex}} \, \varphi_{1} = \mp (\omega_{\mathrm{Ex}} \, \Tilde{\lambda}_{\mathrm{SOT}} + \Tilde{\lambda}_{\mathrm{ME}3}) \, \dot{E}_{y} - \gamma \, \dot{H}_{y},
\end{gathered}    
\end{equation}
\begin{equation}
\begin{gathered}
\label{eq:beta_SODE}
    \Ddot{\beta} + \omega_{\mathrm{A}} \, \omega_{\mathrm{Ex}} \, \beta =  \Tilde{\lambda}_{\mathrm{SOT}} \, \Ddot{E}_{x} - \omega_{\mathrm{A}} \, \Tilde{\lambda}_{\mathrm{ME}3} \, E_{x} \mp \gamma \, \omega_{\mathrm{A}} \, H_{x},
\end{gathered}    
\end{equation}
\begin{equation}
\begin{gathered}
\label{eq:theta_1_SODE}
    \Ddot{\vartheta}_{1} + \omega_{\mathrm{A}} \, \omega_{\mathrm{Ex}} \, \vartheta_{1} =  (\omega_{\mathrm{Ex}} \, \Tilde{\lambda}_{\mathrm{SOT}} + \Tilde{\lambda}_{\mathrm{ME}3}) \, \dot{E}_{x} \pm \gamma \, \dot{H}_{x},
\end{gathered}    
\end{equation}
or using Eq.~\eqref{eq:small_vectors} in the more common form
\begin{equation}
\begin{gathered}
\label{eq:mx_SODE}
    \Ddot{m}_{x} + \omega_{\mathrm{A}} \, \omega_{\mathrm{Ex}} \, m_{x} = \mp \Tilde{\lambda}_{\mathrm{SOT}} \, \Ddot{E}_{x} \pm \omega_{\mathrm{A}} \, \Tilde{\lambda}_{\mathrm{ME}3} \, E_{x} + \gamma \, \omega_{\mathrm{A}} \, H_{x},
\end{gathered}    
\end{equation}
\begin{equation}
\begin{gathered}
\label{eq:my_SODE}
    \Ddot{m}_{y} + \omega_{\mathrm{A}} \, \omega_{\mathrm{Ex}} \, m_{y} = \mp \Tilde{\lambda}_{\mathrm{SOT}} \, \Ddot{E}_{y} \pm \omega_{\mathrm{A}} \, \Tilde{\lambda}_{\mathrm{ME}3} \, E_{y} + \gamma \, \omega_{\mathrm{A}} \, H_{y},
\end{gathered}    
\end{equation}
\begin{equation}
\begin{gathered}
\label{eq:lx_SODE}
    \Ddot{l}_{x} + \omega_{\mathrm{A}} \, \omega_{\mathrm{Ex}} \, l_{x} =  (\omega_{\mathrm{Ex}} \, \Tilde{\lambda}_{\mathrm{SOT}} + \Tilde{\lambda}_{\mathrm{ME}3}) \, \dot{E}_{y} \pm \gamma \, \dot{H}_{y},
\end{gathered}    
\end{equation}
\begin{equation}
\begin{gathered}
\label{eq:ly_SODE}
    \Ddot{l}_{y} + \omega_{\mathrm{A}} \, \omega_{\mathrm{Ex}} \, l_{y} = - (\omega_{\mathrm{Ex}} \, \Tilde{\lambda}_{\mathrm{SOT}} + \Tilde{\lambda}_{\mathrm{ME}3}) \, \dot{E}_{x} \mp \gamma \, \dot{H}_{x}.
\end{gathered}    
\end{equation}


As seen from Eqs.~\eqref{eq:lx_SODE} and~\eqref{eq:ly_SODE}, the linear magnetoelectric effect and SOT enter the above equations for dynamics of the antiferromagnetic vector $\mathbf{l}$ in a similar way and have similar dependencies on spin arrangements in different antiferromagnetic domains.
The \textcolor{newtext}{SOT} at low frequencies was studied, for instance, in metallic antiferromagnets in Ref.~\cite{wadley2016electrical}.
The metallicity is important because the electric field, being screened by free charge carriers, cannot penetrate into the film, but due to large electric conductivity drives the electric current $\mathbf{j}$.
The oscillating electric field can induce displacement current even in materials with poor electric conductivity and thus also in insulating magnets where the screening by free charges is absent and strong electric fields can easily penetrate into the medium. 
Thus, in addition to the THz magnetoelectric effect reported earlier~\cite{bilyk2025control}, the SOT is also likely capable of driving the coherent spin dynamics at THz frequencies.
\textcolor{newtext}{It is worth noting that both discussed torques are field-like.
The physical picture of the new THz effect remains essential the same as the one for N{\'e}el spin-orbit torque in metallic antiferromagnets at zero frequency, where it is also the field-like torque, eventually leads to switching of collinear antiferromagnets rather than damping-like torque~\cite{dal2024antiferromagnetic}.
The electric field $\mathbf{E}^{\mathrm{THz}}$ of a THz pulse generates a THz displacement current $\mathbf{j}_{\mathrm{D}}$, which, via spin-orbit coupling, induces field-like torques that act in opposite directions on the different magnetic sublattices.
This results in spin dynamics that are governed by the antiferromagnetic vector $\mathbf{l}$.
}

\begin{figure}
\centering
\includegraphics[width=1\columnwidth]{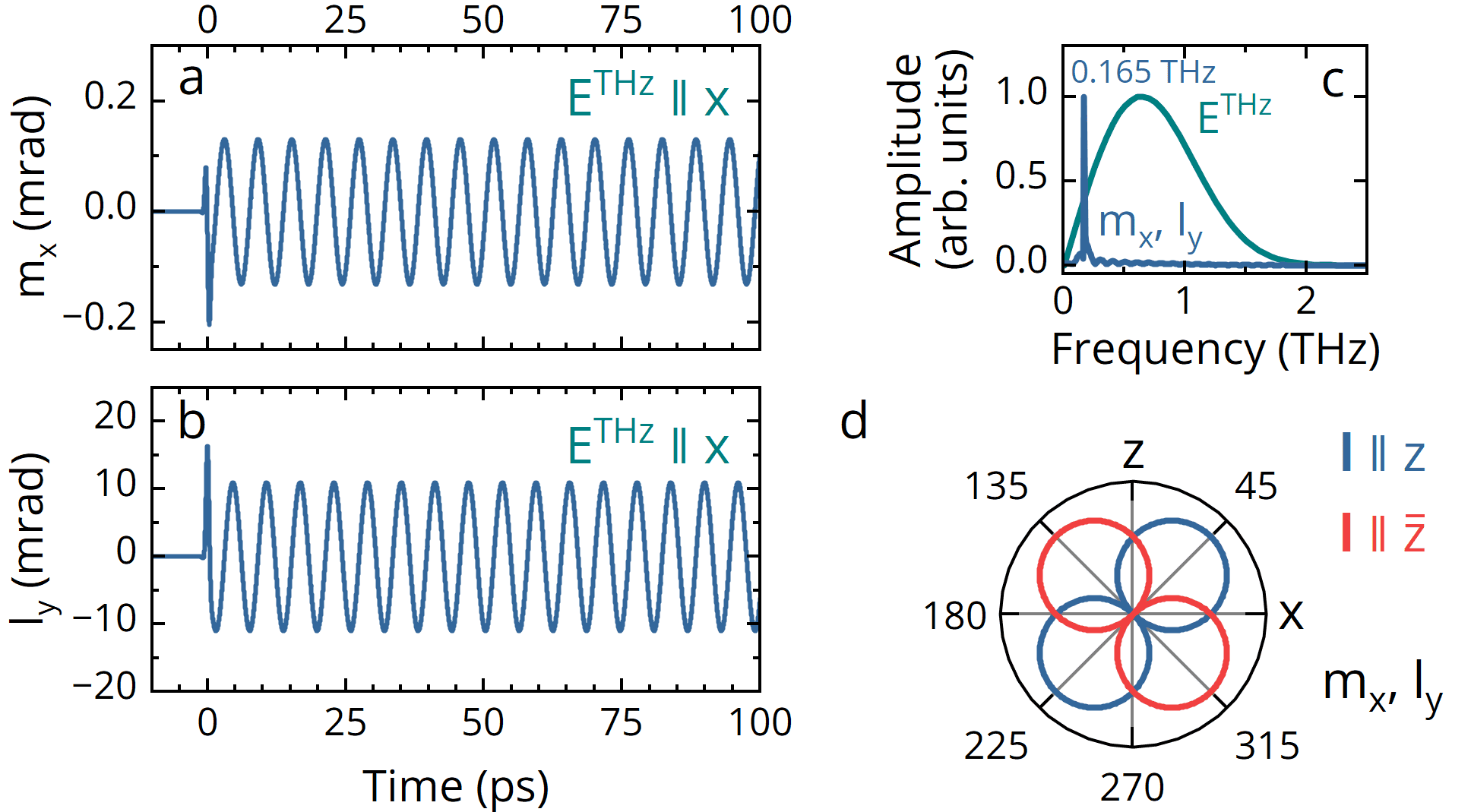}
\caption{\label{fig:spin_dynamics}
Temporal oscillations of (a)~magnetization $m_{x}$ and (b)~antiferromagnetic vector $l_{y}$ components driven by the THz electric field $\mathbf{E}^{\mathrm{THz}} \parallel x$ in a single antiferromagnetic domain of $\mathrm{Cr_{2}}\mathrm{O_{3}}$.
(c)~Normalized Fourier spectra of spin dynamics from (a) and (b) compared to the spectrum of the THz pump pulse.
(d)~Polar diagram of the amplitude of oscillations $m_{x}$ and $l_{y}$ as a function of the THz polarization angle for two opposite antiferromagnetic domain with $\mathbf{l} \parallel z$ (blue) and $\mathbf{l} \parallel \overline{z}$ (red).
}
\end{figure}

\textcolor{newtext}{As experimentally demonstrated in Ref.~\cite{bilyk2025control}, a single-cycle THz pump pulse polarized in the plane of the antiferromagnetic vector $\mathbf{l}$ in \CrO{} induces spin dynamics of comparable amplitude, irrespective of whether the driving force is the THz electric field ($\mathbf{E}^{\mathrm{THz}} \perp \mathbf{l}$) or the THz magnetic field ($\mathbf{H}^{\mathrm{THz}} \perp \mathbf{l}$).
Taking into account that for the electromagnetic plane wave the electric and magnetic fields are equal $|\mathbf{E}^{\mathrm{THz}}| = |\mathbf{H}^{\mathrm{THz}}|$ in Gaussian units, the comparable efficiency in both orthogonal geometries implies that the total electric field torque (magnetoelectric and SOT) is close in absolute value to the Zeeman torque, i.e., the expression $\omega_{\mathrm{Ex}} \, \Tilde{\lambda}_{\mathrm{SOT}} + \Tilde{\lambda}_{\mathrm{ME}3} \simeq -\gamma$ is fair.
Next, we performed simulations of this experiment with the THz pump pulse from Fig.~\ref{fig:displacement_current} polarized in the $xz$ plane, solving Eqs.~\eqref{eq:mx_SODE} and~\eqref{eq:ly_SODE} with the aforementioned torque parameters.
The resulting dynamics of $m_{x}$ and $l_{y}$ for the case $\mathbf{E}^{\mathrm{THz}} \parallel x$ in the single antiferromagnetic domain with $\mathbf{l} \parallel z$ are presented in Figs.~\ref{fig:spin_dynamics}(a) and~\ref{fig:spin_dynamics}(b).
The observed oscillations occur at a frequency of 0.165\,THz, which corresponds to the antiferromagnetic resonance~\cite{foner1963high}, as confirmed by the Fourier spectra in Fig.~\ref{fig:spin_dynamics}(c).
Notably, the oscillation amplitude exhibits a non-trivial dependence on the polarization angle of $\mathbf{E}^{\mathrm{THz}}$, as shown in Fig.~\ref{fig:spin_dynamics}(d).
This behavior stems from the interference between the electric field-induced torques and the magnetic Zeeman torque, consistent with the experimental findings of Ref.~\cite{bilyk2025control}.
Interestingly, this angular dependence undergoes a 90$^{\circ}$ rotation in an antiferromagnetic domain with the opposite orientation of $\mathbf{l}$ [Fig.~\ref{fig:spin_dynamics}(d)].}

Although the available literature lacks data on the magnitude of the SOT contribution for the insulator \CrO, we can estimate it based on known parameters.
First, the static magnetoelectric coefficient $\alpha_{\perp} \simeq - 9 \, \times 10^{-5}$ has been measured in Ref.~\cite{astrov1961magnetoelectric}.
This coefficient is related to the magnetoelectric parameter as $\alpha_{\perp} = \pm \lambda_{\mathrm{ME}3} \, \chi_{\perp}$, where $\chi_{\perp}$ is the perpendicular magnetic susceptibility~\cite{bilyk2025control}.
In the static regime, where SOT effects are negligible, these values yield $\lambda_{\mathrm{ME}3} \simeq - 0.8$~\cite{mukhin1997bwo,foner1963high,astrov1961magnetoelectric}. 
Interestingly, optical measurements reveal a magnetoelectric response of comparable magnitude to the static case~\cite{pisarev1991optical,krichevtsov1993spontaneous,krichevtsov1996magnetoelectric}.
However, in the THz range, the experimental results, in accordance with Eqs.~\eqref{eq:lx_SODE} and~\eqref{eq:ly_SODE}, do not permit the separate determination of the magnetoelectric and SOT contributions.
Instead, only their combined effect $\omega_{\mathrm{Ex}} \, \lambda_{\mathrm{SOT}} \, \pi_{z} \, \varepsilon / (4 \, \pi) + \lambda_{\mathrm{ME}3} \simeq -1$ is observable.
Assuming that the THz magnetoelectric parameter remains close to its static value $\lambda_{\mathrm{ME}3} \simeq - 0.8$, we can estimate that the SOT contribution does not exceed $\lambda_{\mathrm{SOT}} / \omega_{\mathrm{Ex}} \lesssim -0.25$.

\textcolor{newtext2}{
It is worth noting that the amplitude of the spin dynamics drive by the electric field of the THz pulse in \CrO{} [see Figs.~\ref{fig:spin_dynamics}(a) and~\ref{fig:spin_dynamics}(b)] is significantly smaller than in the metallic antiferromagnet $\mathrm{Mn}_{2}\mathrm{Au}$~\cite{behovits2023terahertz,dubrovin2025competition}.
This difference arises because the torque in conducting antiferromagnets such as $\mathrm{Mn}_{2}\mathrm{Au}$~\cite{behovits2023terahertz,dubrovin2025competition} and metallic or semiconducting doped \CrO~\cite{thole2020concepts} is generally larger than in insulating \CrO.
In conducting systems, the torque involve the product of the THz electric field $\mathbf{E}^{\mathrm{THz}}$, the magnitude of which inside the thin film with a thickness less than the penetration depth is units of percent of the incident one, on the very large electrical conductivity $\sigma$.
However, our theory suggests that the SOT in insulator systems can be significantly enhanced at higher frequencies.
Since the displacement current is related to the time derivative of the electric field $\mathbf{j}_{\mathrm{D}} \propto \dot{\mathbf{E}}$, increasing the frequency $\omega$ of the applied field $\mathbf{E}$ leads to a proportional increase in $\mathbf{j}_{\mathrm{D}}$. 
Therefore, we predict that at optical frequencies and above, SOT driven effects in insulators should dominate over those arising from the linear magnetoelectric effect in the linear optics~\cite{krichevtsov1993spontaneous,krichevtsov1996magnetoelectric}, time-resolved magneto-optical experiments~\cite{montazeri2015magneto} and x-ray magnetic linear dichroism~\cite{dc2023observation}. 
This is in stark contrast to the static case, where SOT is zero in insulators unlike metallic antiferromagnets.
}

\section{Conclusions}
\textcolor{newtext}{In summary, we demonstrate that the recently discovered N{\'e}el spin-orbit torque, which has so far been investigated in metallic antiferromagnets at low, near zero frequencies, also becomes accessible in insulating antiferromagnets when the excitation frequency is extended into the THz range.}
Employing symmetry analysis, we reveal a coupling between the antiferromagnetic order parameter and the displacement current induced by the terahertz electric field.
Our analysis shows that this coupling should also include the electric dipole order parameter, which stems from the non-zero dipole moments at the $\mathrm{Cr}^{3+}$ magnetic ion sites.
Using a Lagrangian approach, we derived the equations of spin dynamics when the considered spin-orbit torque competes with the linear magnetoelectric effect, mirroring behavior observed in metallic antiferromagnets.
This indicates that the N{\'e}el spin-orbit torque is a more general magnetic phenomenon, not limited to metallic antiferromagnets.
\textcolor{newtext2}{Crucially, the N{\'e}el spin-orbit torque in insulators is fundamentally different from its metallic counterpart and from the linear magnetoelectric torque, particularly in its dependence on the frequency of the applied electric field.
Indeed, a static electric field cannot induce this torque in insulators.}
\textcolor{newtext2}{In addition, we assume that at THz pumping the N{\'e}el spin-orbit torque alone cannot result in a switching of spins yet.
It does not take away the fact that this mechanism can still be employed in controlling spins in combination with other mechanisms and/or assisting perturbations (such as heat and external static field).
Moreover, we would like to note that switching at THz rates has been a key motivation in antiferromagnetic spintronics, but has not yet been experimentally achieved.
This challenge has prompted the search for mechanisms to excite spins at THz frequencies, particularly THz spin oscillations.}
\textcolor{newtext2}{Nevertheless}, our findings open new avenues for ultrafast spin control in insulating magnetoelectric antiferromagnets by terahertz electric fields, extending beyond the straightforward linear magnetoelectric effect~\cite{bilyk2025control} and sum-frequency excitation~\cite{juraschek2021sum}, \textcolor{newtext2}{and highlighting the underestimated role of insulating materials in THz spintronics}.

\section*{Acknowledgements}
\textcolor{newtext}{We are grateful to M.\,A.~Frolov for fruitful discussions.}
R.\,M.\,D. acknowledges support from Russian Science Foundation Grant No. 24-72-00106.
A.\,K.\,Z. acknowledges support from Russian Science Foundation Grant No. 22-12-00367.
Z.\,V.\,G. acknowledges support from the Ministry of Science and Higher Education of Russia (Agreement No. 075-03-2024-123/1).
A.\,V.\,K. acknowledges support from the European Research Council ERC Grant Agreement No. 101054664 (SPARTACUS).
The authors declare that this work has been published as a result of peer-to-peer scientific collaboration between researchers.
The provided affiliations represent the actual addresses of the authors in agreement with their digital identifier (ORCID) and cannot be considered as a formal collaboration between the aforementioned institutions.

\section*{Data Availability}
The data that support the findings of this article are openly available~\cite{dubrovin_dataset}.

\bibliography{bibliography}

\end{document}